\documentclass[prl,amsmath,amssymb,aps,10pt,superscriptaddress,longbibliography,twocolumn,floatfix]{revtex4-2}
\usepackage{amsmath, amsthm, amssymb, amsfonts,comment}
\usepackage{graphicx} 
\usepackage{siunitx}
\usepackage[colorlinks=true, linkcolor=blue, urlcolor=blue, citecolor=blue, anchorcolor = blue]{hyperref} 
\usepackage{color,xcolor}

\usepackage{dcolumn}
\newcolumntype{d}[1]{D{.}{.}{#1}}
\usepackage[utf8]{inputenc}
\usepackage{natbib}
\usepackage{pdfpages}
\usepackage{pgffor} 
\usepackage{xr} 
\makeatletter
\AtBeginDocument{\let\LS@rot\@undefined}
\makeatother
\def\supplementfilename{supplemental}

\pdfximage{\supplementfilename.pdf}
\def\numbersupplementpages{\the\pdflastximagepages}

\newif\ifarXiv
\arXivtrue

\usepackage{soul} 
\usepackage[normalem]{ulem}

\DeclareSIUnit{\sqrthz}{\ensuremath{\sqrt{\text{\hertz}}}}

\definecolor{orange}{rgb}{0.99,0.5,0.31}

\newcommand{\bET}{\beta^T_{\rm E}} 
\newcommand{\bRFT}{\beta^T_{\rm RF}}
\newcommand{\bE}{\beta_{\rm E}} 
\newcommand{\bRF}{\beta_{\rm RF}}

\newcommand{\dP}{\delta_{\rm P}}
\newcommand{\dC}{\delta_{\rm C}}
\newcommand{\dRF}{\delta_{\rm RF}}
\newcommand{\keff}{k_{\rm eff}}
\newcommand{\kP}{k_{\rm P}}
\newcommand{\kC}{k_{\rm C}}
\newcommand{\CC}{C_{\rm C}}
\newcommand{\CRF}{C_{\rm RF}}
\newcommand{\omgC}{\Omega_{\rm C}}
\newcommand{\omgRF}{\Omega_{\rm RF}}

\newcommand{\IET}{I^T_{\rm E}}
\newcommand{\IRFT}{I^T_{\rm RF}}
\newcommand{\IPT}{I^T_{\rm P}}
\newcommand{\lP}{\lambda_{\rm P}}
\newcommand{\lC}{\lambda_{\rm C}}

\newcommand{\IEo}{I^0_{\rm E}}
\newcommand{\IRFo}{I^0_{\rm RF}}
\newcommand{\IPo}{I^0_{\rm P}}
\newcommand{\IE}{I_{\rm E}}
\newcommand{\IRF}{I_{\rm RF}}
\newcommand{\IP}{I_{\rm P}}

\newcommand{\TBB}{T_{\rm BB}}

\newlength{\lengthofminus}
\newlength{\lengthofone}
\begin{document}

\title{The Impact of Thermal Fields on Rydberg Atom Radio Frequency Sensors}

\date{\today}

\author{Channprit Kaur}
\affiliation{ 
National Research Council Canada, Metrology, Ottawa, Ontario K1A 0R6, Canada
}
\author{Pinrui Shen}
\affiliation{ Quantum Valley Ideas Laboratories, 485 Wes Graham Way, Waterloo, Ontario N2L 6R1, Canada}
\author{Donald Booth}
\affiliation{ Quantum Valley Ideas Laboratories, 485 Wes Graham Way, Waterloo, Ontario N2L 6R1, Canada}

\author{Andrew Todd}
\affiliation{ 
National Research Council Canada, Metrology, Ottawa, Ontario K1A 0R6, Canada
}
\author{James P. Shaffer}
\email{e-mail: jshaffer@qvil.ca}
\affiliation{ Quantum Valley Ideas Laboratories, 485 Wes Graham Way, Waterloo, Ontario N2L 6R1, Canada}

\begin{abstract}
Rydberg atom radio frequency sensors are unique in a number of ways, including possessing extraordinary carrier bandwidth, self-calibration and accuracy. In this paper, we examine the impact of thermal radiation on Rydberg atom sensors. Antennas are limited by their thermal background, while Rydberg atom sensors are coherent sensors. Incoherent thermal radiation does not limit Rydberg atom sensors in the same way as an antenna. The primary consequence of a thermal radiation field on Rydberg atom sensors is to decrease their coherence, as the decay rates of the Rydberg states used for sensing the radio frequency field are increased due to the thermal field, i.e. blackbody, modification of the atomic decay rates. Thermal and coherent field excitation are fundamentally different in that thermal fields produce statistically independent excitations with well-defined frequency, polarization, and propagation direction, while coherent states are coherent superpositions of photon number states. Consequently, thermal fields do not contribute to the coherences of the density matrix that are used for Rydberg atom sensing, except for damping them.

\end{abstract}
\maketitle

Rydberg atom-based radio frequency (RF) sensors have many advantages over traditional RF technology, such as possessing extraordinary carrier bandwidth, an intrinsic calibration and higher accuracy \cite{sedlacek2012microwave, gordon2019weak, fan2015atom, schlossberger2024rydberg, PhysRevApplied.20.L061004}. The readout and preparation scheme for Rydberg atom sensors uses electromagnetically induced transparency (EIT) or absorption (EIA). These processes are coherent, they exhibit interference amongst different excitation pathways within the atom. In Refs.$\,$\cite{schmidt2024rydberg, schmidt2025all}, it was shown that Rydberg atom sensors operating in the weak probe limit depend only on the coherences, i.e. non-secular terms of the density matrix, and their decay rates that are involved in the interfering transition pathways.

Rydberg atom sensors are fundamentally limited by photon and atomic shot noise \cite{fan2015atom,Kumar2017, bussey2022}, whereas traditional receivers are limited by thermal noise \cite{HANSEN200232-1}. Shot noise and projection noise can be reduced using suitable geometry, suggesting that Rydberg atom sensors can be more sensitive than conventional receivers. It has not yet been shown that Rydberg atom sensors can be more sensitive than state-of-the-art receivers. A necessary step for investigating if Rydberg atom sensors can be more sensitive than conventional receivers, given the difference in the origin of their noise limitations, is to understand how thermal fields impact the Rydberg atom sensing process. In this paper, we show that thermal fields reduce the sensitivity of Rydberg atom sensors through modification of the decay rates of the Rydberg states used for the sensor. A thermal field does not influence the EIT sensing process by acting as a competing noise field on the target Rydberg level transition. Experiments are described and data presented that confirm the theory presented in Refs.$\,$\cite{schmidt2024rydberg, schmidt2025all}, indicating that black-body radiation (BBR), a thermal field, affects Rydberg atom sensors through increasing the decay rate of the relevant atomic coherences.

\begin{figure}
     \centering
     \includegraphics[width=1\linewidth]{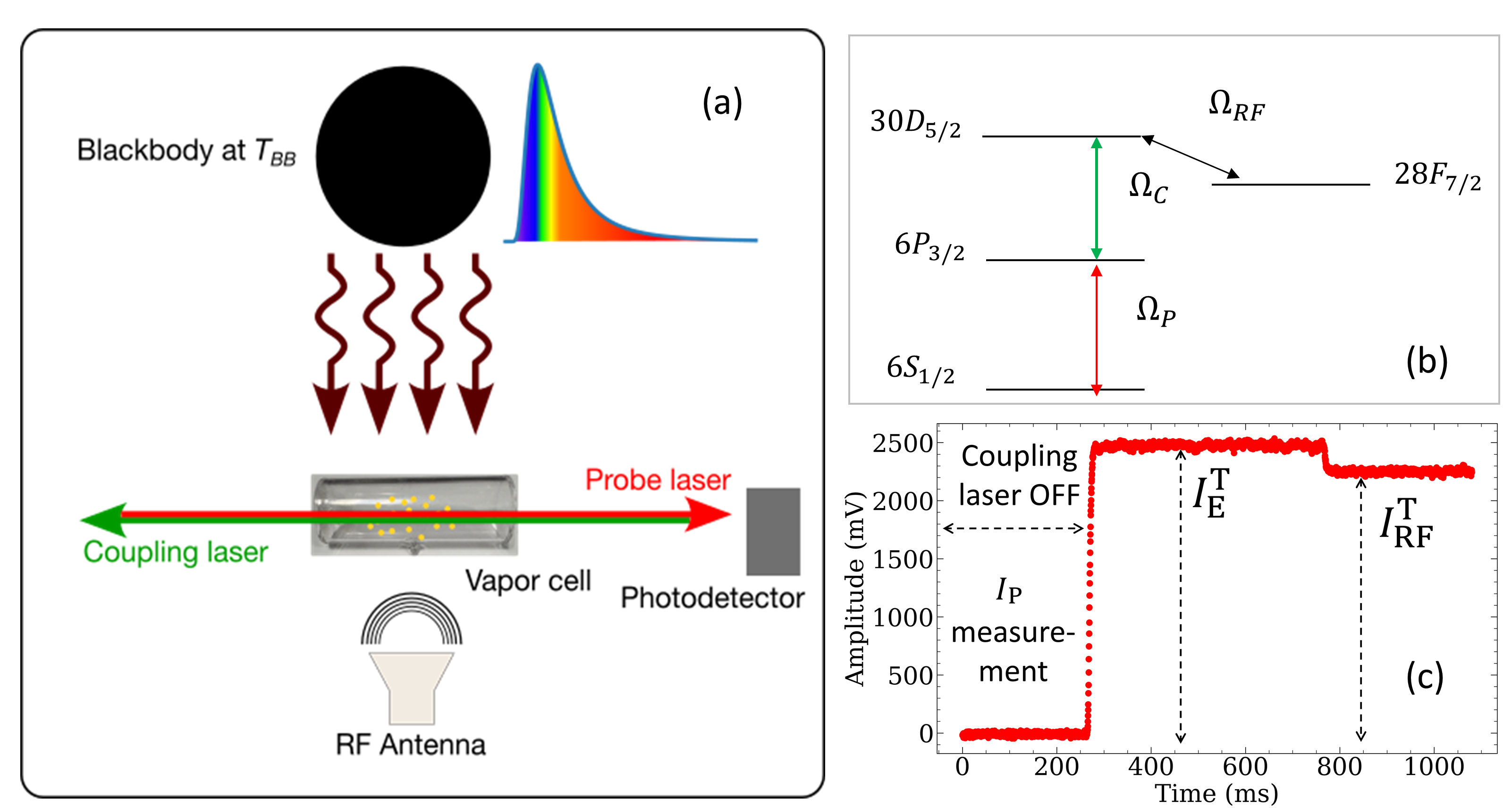}
     \caption{(a) Experimental setup. (b) Energy level diagram employed. The probe laser was locked to the Cs $|6S_{1/2},F=4\rangle$ to $|6P_{3/2},F=5\rangle$ transition. The coupling laser was tuned to the $|6P_{3/2},F=5\rangle$ to $30D_{5/2}$ transition  while the RF field drove the $30D_{5/2}$ to $ 28F_{7/2}$ transition. (c) Time-domain trace of the lock-in amplifier output signal where $\IET$ and $\IRFT$ are the EIT transmission intensities with and without RF, respectively.}
     \label{fig:1}
 \end{figure}

Several authors have postulated the influence of BBR on Rydberg atom sensors using an analogy with conventional antennas, i.e. using the concept of a noise temperature \cite{chen2025, santamariabotello2022comparisonnoisetemperaturerydbergatom, simons2018electromagnetically}. To our knowledge, no one has yet investigated how a thermal field modifies the Rydberg atom sensing process from an atomic perspective. The concept of thermal noise temperature as applied to antennas does not work for Rydberg atom sensors\cite{supp}.

It is well-known that thermal fields and coherent fields act on atoms and molecules differently, even at the single excitation limit. Fundamental work has elucidated the differences in large molecules, whose natural photochemistry is determined by interaction with sun- and moonlight, rather than the coherent laser pulses commonly used to investigate their dynamics \cite{PhysRevA.87.022106, tscherbul2014coherent,BrumerShapiro12}. Thermal fields produce statistically independent excitations with well-defined energy, polarization and propagation direction. Coherent states are formed from coherent superpositions of photon number states. BBR is characterized by a diagonal density matrix because there is no phase relationship between different components of the mixture. Typical coherence times for BBR, e.g. the sun, are on the order of 10's of fs \cite{Mehta1964-I, Mehta1964-II}, much faster than any signal detectable by a Rydberg atom sensor given their baseband bandwidths. BBR can be described using rate equations which modify the decay rates of the atoms \cite{Gallagher1979,cooke1980,HAROCHE1985}, and shift the atomic energy levels \cite{Hollberg1984, ovsiannikov}.  BBR cannot induce coherences in the EIT or EIA process. Instead, BBR disturbs the sensor by changing the coherence parameters through the damping of the atomic density matrix coherences. In contrast, thermal fields induce dissipative noise currents in antennas. 

Thermal fields can produce coherences in exceptional cases, but these are not relevant for Rydberg atom sensors \cite{Mehta1964-I, Mehta1964-II,PhysRevA.87.022106}. As a thermal field turns-on or is perturbed, coherences can be produced due to the time-dependent envelope of the electromagnetic field. The coherences generated are on the timescale of the turn-on. Likewise a thermal pulse can generate coherences. The mechanism for these instances is the excitation of a wavepacket consisting of degenerate or nearly degenerate states. Starting from an initial Rydberg state, a thermal field can excite, either through stimulated emission or absorption, a wavepacket in the Rydberg manifold comprising degenerate states coupled by the same excitation. The coherences generated are between the superposed excited states, therefore not involved in the Rydberg sensing process \cite{tscherbul2014coherent}. 

\begin{figure}
    \centering
    \includegraphics[width=1\linewidth]{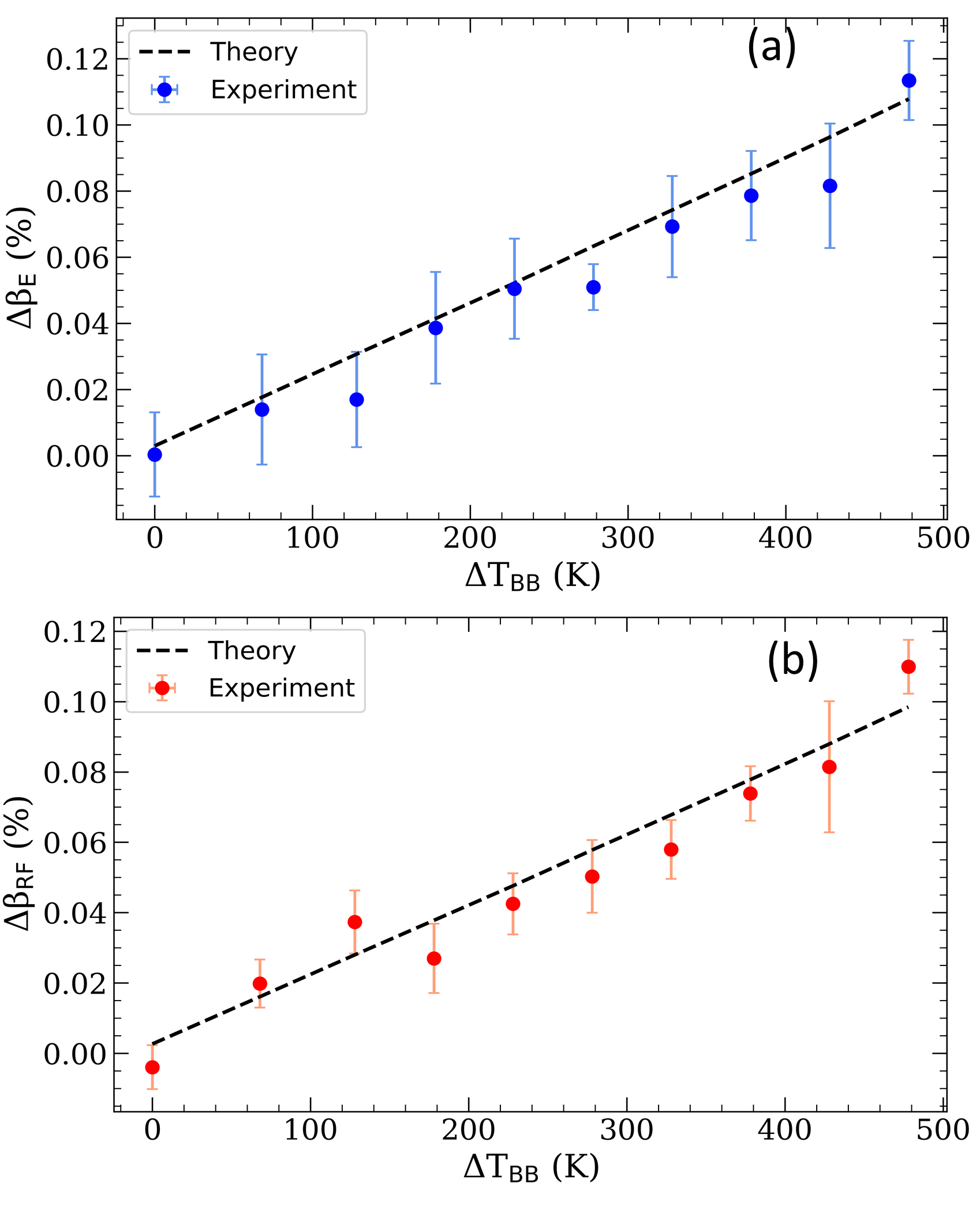}
    \caption{(a) $\Delta \bE$ as a function of $\Delta \TBB=\TBB-T_{\rm room}$, where $\TBB$ is the BBR temperature and $T_{\rm room}$ is the room temperature. (b) $\Delta \bRF$ as a function of $\Delta \TBB=\TBB-T_{\rm room}$. Each data point is the average over 160 measurements. The error bars are the standard deviation at each measurement point. The error on the temperature axis are $\pm0.4~\rm K$, too small to be visible. The dashed lines are the numerically calculated results.}
    \label{fig:2}
\end{figure}

In the weak probe regime, the transmission of the probe laser beam is described by the Beer-Lambert law, $\IP =I_0\exp(-\beta z)$, where $I_0$ is the initial intensity of the probe laser and $\IP$ is the intensity that is detected after passing through the sample and z is the length of vapor cell. The absorption coefficient, $\beta$, describes the attenuation of the probe laser due to spontaneous emission, assuming $\beta$ real. In the two-photon EIT scheme with an RF field, FIG.~\ref{fig:1}, the absorption coefficient can be written as \cite{schmidt2024rydberg},
\begin{align}
\bRF = A \kP \rm Im \int_{-\infty}^{+\infty} &\frac{1}{2\dP(v) - j} \frac{1}{1 - \frac{\frac{\CC^2}{(2\dP(v) - j)(2\dC(v) - j)}}{1 - \frac{\CRF^2}{(2\dP(v) - j)(2\dRF(v) - j)}}} \notag \\
&\times P(v) \, dv,
\label{eq:1}
\end{align}
where $P(v)$ is the Maxwell-Boltzmann distribution for the thermal Cs gas,
\begin{equation}
    P(v)=\sqrt{\frac{m}{2\pi k T_{\rm Cs}}}exp\left(-\frac{mv^2}{2k T_{\rm Cs}}\right)=\frac{1}{\sqrt\pi\bar{v}}exp\left(\frac{-v^2}{{\bar{v}}^2}\right).
    \label{eq:2}
\end{equation}
$m$ is the atomic mass, $k$ is the Boltzmann constant, $v$ is the atomic velocity and $T_{\rm Cs}$ is the temperature of the Cs gas. $\bar{v}= \sqrt\frac{2k T_{\rm Cs}}{m}$ is the most probable velocity of the atoms. The prefactor $A = 3\rho\lP^3/16\pi^2$ depends on the Cs atomic density, $\rho$, and probe laser wavelength, $\lP$. 

\begin{figure}
    \centering
    \includegraphics[width=1\linewidth]{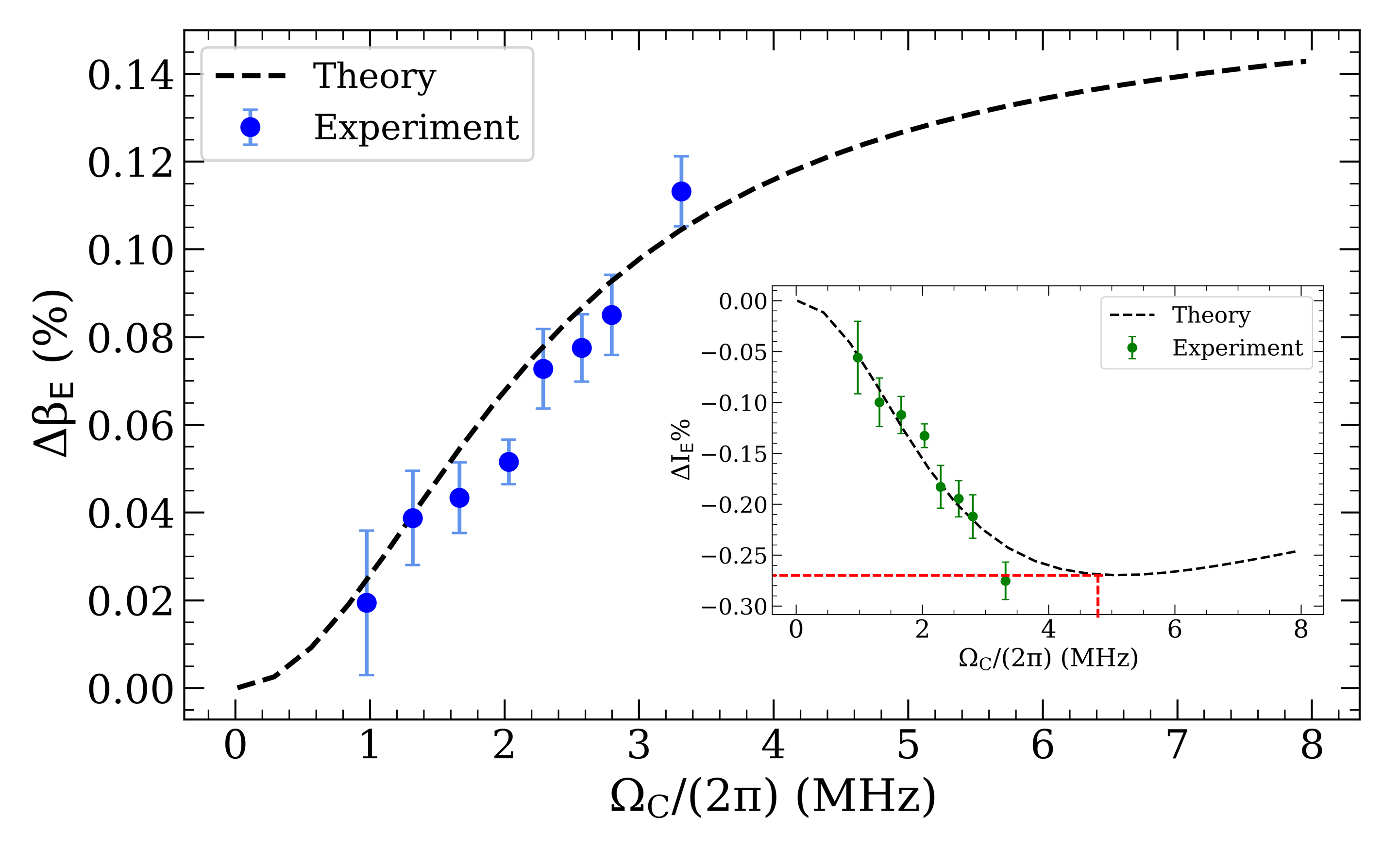}
    \caption{Relative change in absorption coefficient, $\bE$, as a function of the coupling laser Rabi frequencies at $T_{\rm BB}= 773.15$~K. The inset shows the relative change in EIT amplitude, $\IE$, as a function of coupling laser Rabi frequency. The largest change in EIT amplitude happens when $\omgC=4.5 \rm MHz$, as indicated by the red dashed line in the inset. The error bars in both plots are reflective of the standard deviation of each measurement point.}
    \label{fig:3}
\end{figure}

In the case where all fields are on resonance, Doppler-induced detunings in Eq.~\ref{eq:1} are defined as $\dP = \kP \bar{v}/\gamma_2$, $\dC = \keff \bar{v}/\gamma_3$, and $\dRF = \keff \bar{v}/\gamma_4$, where $\keff = \kP - \kC$ accounts for wavevector mismatch in a counter-propagating geometry. $\kP=2\pi/\lP$ and $\kC=2\pi/\lC$ are the wave vectors of the probe and coupling lasers, respectively. In our experiments, the wavelength of the probe laser and the coupling laser are precisely obtained by locking the lasers to a frequency comb, $\lP = 852.356382791~ \rm~nm$, and $\lC = 512.048504537~ \rm~nm$.

The atom-field interaction coherence parameters (coherence parameter), $\CC=\omgC/\sqrt{\gamma_2\gamma_3}$ and $\CRF=\omgRF/\sqrt{\gamma_3\gamma_4}$, represent the degree of coherence of the coupling laser field and the RF field, respectively\cite{schmidt2024rydberg,schmidt2025all}. $\omgC$ and $\omgRF$ are the corresponding Rabi frequencies. $\omgC = 2\pi \times 3.5\,$MHz and $\omgRF = 2\pi \times 600\,$kHz are determined by measurements. The EIT absorption coefficient in the absence of an RF field, $\bE$, can be obtained by evaluating Eq.~\ref{eq:1} at $\CRF=0$. $\gamma_2$, $\gamma_3$ and $\gamma_4$ in the coherence parameters are the total dephasing rates of the Cs $6P_{3/2}$, $30D_{5/2}$ and $28F_{7/2}$ states, respectively. 

The experimental setup shown in FIG.~\ref{fig:1} is designed to measure the probe laser absorption coefficient in the presence of BBR. A cylindrical Cs vapor cell with a diameter of $25.4\,$mm and length of $71.8\,$mm is held at room temperature. An EIT signal from the vapor cell is generated using spatially overlapped, counter-propagating probe ($\sim852$~nm) and coupling ($\sim509$~nm) laser beams with $\sim 200\rm~\mu \textrm{m}$ and $\sim 190\rm~\mu\textrm{m}$
$1/e^2$ radii, respectively. For the measurements, the power of the probe laser is $\sim 150\rm~nW$ so that the experiment operates in the weak probe regime. The power of the coupling laser is $\sim 80\rm~mW$. The lasers are intensity stabilized using liquid crystal amplitude stabilizers which keep the power within $<0.5\%$ of the set value. The lasers are tuned to resonance with the transitions of the ladder system shown in FIG.~\ref{fig:1}. The lasers are linearly polarized and parallel to each other. Both lasers are locked to a frequency comb stabilized to a GPS steered rubidium clock, resulting in laser line widths of less than 200~Hz and a fractional frequency stability of $1.2 \times 10^{-12}$ for a $10\,$s averaging time.

The experiment starts by turning on the probe laser for $300\,$ms to measure its transmission, $\IPo$, through the vapor cell using an avalanche photo-diode detector. The transmission of the probe laser is used to determine $\rho$ \cite{supp}. $\rho \sim 9.6 \times 10^{15}$ m$^{-3}$ for the measurements presented in the manuscript. After determining $\rho$, a vapor pressure model \cite{Alcock1984} is used to derive the temperature of the Cs atomic gas $T_{\rm Cs}$ and the most probable velocity $\bar{v}$, used for $P(v)$. Variations in $\rho$, $T_{\rm Cs}$, and $\bar{v}$ were accounted for at each BBR temperature.   Next, the coupling laser is applied. It is modulated by an acousto-optical modulator at $50\rm~kHz$. The probe laser transmission, $\IEo$, under EIT conditions is measured using a lock-in amplifier. After a period of $300\,$ms, an RF field is applied via a waveguide placed $\sim 15$~cm above the vapor cell. The frequency of the RF field is set to $133.344\,$GHz, which couples $30D_{5/2}$ and $28F_{7/2}$. The RF field strength is set to $<0.1~\textrm{V}\,\textrm{m}^{-1}$ so that the experiment operates in the amplitude regime \cite{Kumar2017}, where the EIT peak splitting is not discernible. The RF field Rabi frequency is measured to be $\omgRF= 2\pi\times 600\rm~kHz$. The probe transmission with the RF on, $\IRFo$, is also measured for $300\,$ms. $\IEo$ and  $\IRFo$ are used to determine the change in probe laser transmission due to the RF field at room temperature, $T_{\rm room}$. Next, the BBR is applied to the vapor cell by opening a motorized shutter in front of a flat-plate BBR source located $\sim$40~cm from the vapor cell. The temperature of the BBR source can operate between room temperature and $773.15\rm~K$. The measurements of the probe laser transmission were repeated with the BBR illuminating the vapor cell for probe only, $\IPT$; EIT, $\IET$; and EIT with the RF field, $\IRFT$, using the same averaging times and apparatus as the BBR source-free measurements. After these measurements, the motorized shutter is closed. One complete measurement cycle takes $2\,$s. The duration is chosen to obtain sufficient averaging without significantly radiatively heating the vapor cell. For each BBR temperature, the experimental cycle is repeated 160 times.

To determine the impact of the BBR, differential signals were calculated. The relative change in $\IE$ and $\IRF$ is calculated as $\Delta I_{\rm E} = (\IET-\IEo)/\IEo$ and $\Delta I_{\rm_{RF}}= (\IRFT-\IRFo)/\IRFo$, respectively.
The change in absorption coefficients, $\Delta \beta_{\rm E}$ and $\Delta \beta_{\rm RF}$ are obtained from the measured intensities using the Beer-Lambert law. $\Delta \beta_{\rm E}$ and $\Delta \beta_{\rm RF}$ as a function of BBR temperature are presented in FIG.~\ref{fig:2}.  An approximate increase of $0.12\%$ is observed in $\Delta \beta_{\rm E}$ and $\Delta \beta_{\rm RF}$ corresponding to a $0.3\%$ decrease in both $\Delta I_{\rm E}$ and $\Delta I_{\rm RF}$ as the BBR temperature ranges from room temperature to $773.15\,$K.

The BBR modifies the absorption coefficients, $\bE$ and $\bRF$, by changing the decay rates of the Rydberg states, $\gamma_3$ and $\gamma_4$ which influence the coherence parameter, Eq.~\ref{eq:1}. The ground state, $6S_{1/2}$ and the intermediate excited state, $6P_{3/2}$, are relatively insensitive to thermal radiation, so their interaction with the BBR can be ignored. The Rydberg states possess large dipole matrix elements, enabling strong coupling to BBR.

The BBR induced decay rates are given by \cite{Gallagher1979,cooke1980effects},
\begin{equation}
\gamma_{\rm BB} \approx \frac{4\alpha^3k\TBB}{3n_{\rm eff}^2},\
\label{eq:3}
\end{equation}
where $\alpha$ is the fine structure constant, $\TBB$ is BBR temperature. The dependence on the quantum number as well as the angular momentum is encoded in the effective principal quantum number, $n_{\rm eff}$,  which is a function of the principal quantum number, $n$,  the angular momentum, $l$, and the total angular momentum, $J$. For the same $n$, the BBR induced decay rate is lower generally for higher $l$. Semi-empirical expressions for BBR-induced depopulation in high-$n$ Rydberg states ($n \gg 80$) have also been provided by Beterov et al.\cite{beterov2009quasiclassical}, using a quasiclassical approximation. For our work we obtained BBR induced decay rates using ARC \cite{vsibalic2017arc} which agrees well with the semi-empirical expressions in Ref.~\cite{beterov2009quasiclassical}. For the $30D_{5/2}$ and $28F_{7/2}$ Rydberg states, with the BBR at $773.15\,$K, the decay rates are 70\% and 50\% higher than the spontaneous decay rate respectively. The contribution of other decay mechanisms such as collisional and transit time broadening to $\gamma_3$ and $\gamma_4$ are discussed in Ref.~\cite{supp}.

Using Eq.~\ref{eq:1}, the absorption coefficients, $\bET$ and $\bRFT$ at each BB temperature were numerically calculated. The relative changes in the absorption coefficients with the RF field, $\Delta \bRF$, and without the RF field, $\Delta \bE$, were obtained and are presented in FIG.~\ref{fig:2}. They agree, within experimental error, with the experimental results, demonstrating that the changes in the absorption with $\TBB$ are induced by BBR enhanced decay rates. The BBR incoherently drives the decay rates of the $28F_{7/2}$ and $30D_{5/2}$ Rydberg states, reducing the coherence parameters of the coupling laser and RF fields. 

The single BBR photon Rabi frequency is less than 40~Hz. The coherence time at $300\,$K is around $2 \times 10^{-14}\,$s. The coherence parameter for the BBR field is vanishingly small. The BBR field cannot induce meaningful changes to the EIT signal by driving the Rydberg coherence. The observed changes in FIG.~\ref{fig:2} are dominated by the BBR incoherent process.

The BBR induced change in the absorption coefficient of the probe laser in the EIT scheme can also be varied by changing the coupling laser's Rabi frequency according to Eq.~\ref{eq:3}. We measured the BBR effects at different $\omgC$ with the BBR temperature fixed at $773.15\rm~K$. The measurements were compared with the numerical calculations at different $\omgC$, which are presented in FIG.~\ref{fig:3}. Within the experimentally achievable range of $\omgC$, the experimental and theoretical results agree, within experimental error, providing further evidence that the equations in \cite{schmidt2024rydberg,schmidt2025all} are accurate for describing the experimental data.

The sensitivity of Rydberg atom sensors is impacted by a thermal field. In the weak probe and large coupling regime, the absorption coefficient can be written as \cite{schmidt2024rydberg},
\begin{equation}
\bRF=Ak_{\rm p}\frac{\gamma_2}{\gamma_4(T)\omgC^2}\omgRF^2+\bE=S(T)\omgRF^2+\bE.
\label{bbrest}
\end{equation}
$S(T)$ and the transition dipole moment of the RF sensing transition determine the sensitivity. An increase in the BBR temperature increases the BBR induced decay rate, $\gamma_4(T)$, which reduces the RF sensitivity. Although Eq.~\ref{bbrest} is applicable to a Doppler free sensor, it can be used to estimate the change in the minimum detectable RF field due to the BBR \cite{supp} at the probe laser shot noise limit. Analytic expressions similar to Eq.~\ref{bbrest} for Doppler limited and wave-vector matched conditions can be found in Ref.~\cite{schmidt2024rydberg}. Likewise, Eq.~\ref{eq:1} can be used for numerical calculations.

We use Eq.~\ref{eq:1} to numerically calculate the effect of the BBR here. For the RF transition between $55D_{5/2}$ and $53F_{7/2}$ with frequency $19.4\,$GHz, as an example, the minimum detectable RF field at $\TBB=0\rm~K $ is $E_{\rm min}=\ 19.04 \rm~\mu V/cm$. The minimum detectable RF field increases to $E_{\rm min}=\ 19.05\rm~\mu V/cm$ at $\TBB=296.15\rm~K $ and $E_{\rm min}=\ 19.06 \rm~\mu V/cm$ at $\TBB=773.15\rm~K $. Therefore the RF sensitivity decreased by 0.05~\% and 0.1~\% due to the thermal field, $\TBB=296.15 \rm~ K$ and $\TBB=773.15 \rm~ K$ respectively\cite{supp}. For these estimates, we used typical values for Rydberg atom sensors. The probe laser power was taken to be $1\,\mu$W, a vapor cell length of $71.8~\rm mm$ and $\omgC$ was $4.5\,$MHz. A linear approximation for the probe laser absorption was applied for the calculations.

The advantageous feature of the insensitivity to thermal fields is that coherent fields can, in principle, be detected below the thermal limit of an antenna. While the minimum field estimates are not below the thermal limit of an antenna, they can be coupled with resonators such as that found in Ref.~\cite{Amarloo25}. The enhancement from such a device can put the sensitivity well-below the thermal limit of an antenna.

The reduction in sensitivity originates from the incoherence properties of the thermal field which broaden the EIT transmission feature. The thermal field affects the sensor only through this mechanism. Incoherent fields are difficult to detect due to their statistical properties. Rydberg atom sensors are not ideal for measurements of fields with widely fluctuating phase. Thermal fields, i.e. radiometry, and some types of electromagnetic interference (EMI), so-called broadband EMI, for example, will only manifest through the decay rates. The measure of how much fluctuation can be tolerated is also related to the bandwidth of the Rydberg sensor through the time-frequency relationship.

We have experimentally demonstrated and explained the impact of incoherent thermal radiation on Rydberg RF sensors. While the target coherent field drives well-defined quantum dynamics, e.g., Rabi oscillations, thermal fields cause population redistribution which affects the sensor by increasing the decay rates of the coherences involved in the optical detection and readout. This will reduce the sensitivity of the Rydberg RF sensor originates from Rydberg transition broadening. We have found the sensitivity decreases as the BB temperature increases and formulated the RF sensitivity as a function of the black body temperature and found the sensitivity is reduced by 0.1~\% at $T_{\rm BB}=773.15 \rm~ K$.

C.K. and A.T. acknowledge  support from QRDI, National Research Council Canada. P.S., D.B. and J.S. acknowledge support from National Research Council Canada IoT: QSC (QSP-058-1 and QSP-105-1) and from DARPA (HR0011-21-C-0141). Any opinions, findings and conclusions or recommendations expressed are those of the author(s) and do not necessarily reflect the views of DARPA.

%
\clearpage
\ifarXiv
    

\section*{Supplementary material}

# Supplementary Material: The Impact of Thermal Fields on Rydberg Atom Radio Frequency Sensors

Channprit Kaur,[1] Pinrui Shen,[2] Donald Booth,[2] Andrew Todd,[1] and James P. Shaffer[2, *]

[1] National Research Council Canada, Metrology, Ottawa, Ontario K1A 0R6, Canada

[2] Quantum Valley Ideas Laboratories, 485 Wes Graham Way, Waterloo, Ontario N2L 6R1, Canada

(Dated: August 24, 2025)

## I. Measurement of Cs number density from probe laser transmission

The transmission of the probe laser in the weak-probe regime can be described by the Beer–Lambert law. This law relates the absorption to the propagation distance, $z$, through the atomic medium as

$$I_p^{exp}(z) = I_0 e^{-\beta z} \tag{1}$$

where $I_0$ is the initial probe laser intensity and $I_p^{exp}$ is the detected probe intensity after the Cs vapor cell. Both $I_0$ and $I_p^{exp}$ are determined experimentally. The absorption coefficient, $\beta$, can be written as [1],

$$\beta = \frac{3\pi\rho(T_{Cs})\lambda_p^3}{16\pi^3} k_P \frac{\gamma_2}{\Omega_P} \int_{-\infty}^{+\infty} P(v) \frac{1}{1 + 4\delta_p} dv, \tag{2}$$

where $\rho(T_{Cs})$ is the Cs atomic density and can be determined from the vapor pressure model [2] for a given temperature of the Cs gas, $T_{Cs}$. $\gamma_2 = 2\pi \times 5.234(13)$ MHz [3] is the spontaneous emission rate of the probe transition for the $|6P_{3/2}\rangle$ state, $\lambda_p = 852.356382791$ nm is the probe laser wavelength, $k_P$ is the probe laser wavevector, $\Omega_p = 2\pi \times 500$ kHz is the probe laser Rabi frequency and $\delta_p = k_P \bar{v}/\gamma_2$ is the probe laser detuning.

For a gas of moving atoms, the thermal average velocity distribution, $P(v)$, is given by

$$P(v) = \sqrt{\frac{m}{2\pi k T_{Cs}}} \exp\left(-\frac{mv^2}{2kT_{Cs}}\right) = \frac{1}{\sqrt{\pi}\bar{v}} \exp\left(\frac{-v^2}{\bar{v}^2}\right). \tag{3}$$

Here $m$ is the mass of the atoms, $\bar{v} = \sqrt{\frac{2kT_{Cs}}{m}}$ is the most probable velocity of Cs atoms, $k$ is the Boltzmann constant. We used the nonlinear least squares fitting method to fit experimentally measured $I_0$ and $I_p^{exp}$ values to Eq. 1 with the single fitting parameter, $T_{Cs}$. The extracted value $T_{Cs}$ can then be used to calculate the atomic density of Cs gas, $\rho(T_{Cs})$ using the vapor pressure model [2] and the ideal gas law, $P = \rho k T_{Cs}$.

## II. Decay processes and their effect

In addition to spontaneous decay, mechanisms such as blackbody(BB) induced decay rates, atomic collisions, and transit effects also contribute to the overall coherence loss. In this section, we will explore the dominant

decay process affecting our experimental system. The total decay rates for the states Cs $|6P_{3/2}\rangle$, $|30D_{5/2}\rangle$ and $|28F_{7/2}\rangle$ are denoted by $\gamma_2$, $\gamma_3$ and $\gamma_4$ respectively. For the intermediate excited state, $6P_{3/2}$, there are three dominated decay mechanisms: the spontaneous decay $\gamma_{2,spon} = 2\pi \times 5.234(13)$ MHz [3], the transit induced decay, $\gamma_{2,trans} \sim \bar{v}/d \approx 2\pi \times$ 80 kHz, due to the finite interaction time of Rydberg atoms in laser field, where $d$ is the laser beam diameter. The third one is collisional decay with the ground state Cs atoms, which can be modeled as, $\gamma_{2,col} = \rho\sigma_g\bar{v}$. Here the collision cross section $\sigma_g = 1.306 \times 10^{-16}$ m$^2$, which is estimated using the geometrical cross section[4],

$$\sigma_g = \pi a_0(n^2/2)[5n^2 + 1 - 3l(l+1)]\sqrt{l(l+1)}/(2l+1), \tag{4}$$

where $l$ is the angular quantum number and $a_0$ is Bohr's radius.

For the decay rates of the Rydberg states, $\gamma_3$ and $\gamma_4$, the spontaneous decay rate is calculated using ARC (Alkali.ne Rydberg Calculator) [5] and the transit induced decay is calculated using the same method as the $6P_{3/2}$ state. The ground state atom-Rydberg state atom collisional decay rate can be expressed as, $\gamma_{G-R} = \rho\sigma_{G-R}\bar{v}$. The ground-Rydberg state collision cross section, $\sigma_{G-R}$ can also be calculated using the geometrical cross-section model, Eq. 4. The values of $\sigma_{G-R}$ for $30D_{5/2}$ and Cs $28F_{7/2}$ states are $8.69 \times 10^{-15}$ m$^2$ and $6.63 \times 10^{-15}$ m$^2$ respectively.

In addition, the Rydberg-Rydberg collisional decay rate, $\gamma_{R-R} = \rho_R\sigma_{R-R}\bar{v}$, where $\rho_R$ is the density of Cs atoms in Rydberg states, which is about a factor of $4 \times 10^{-4}$ of the Cs ground state atoms density in our case. The Rydberg-Rydberg collision cross-section is calculated using the experimentally measured equation in Ref. [6], $\sigma_{R-R} = 2.37 \times 10^{-18} n_{eff}^4$ m$^2$. For Cs $30D_{5/2}$ and $28F_{7/2}$ states, the values of $\sigma_{R-R}$ are $1.92 \times 10^{-12}$ m$^2$ and $1.456 \times 10^{-12}$ m$^2$, respectively. Due to the low density of the Rydberg atoms and their large interatomic spacing, the Rydberg-Rydberg blockade induced decay rate is negligible.

Finally, the decay rate induced by blackbody radiation, $\gamma_{BB}$, can modeled by the following relation [7]

$$\gamma_{BB} \approx \frac{4\alpha^3 k T_{BB}}{3n_{eff}^2} \tag{5}$$

where $\alpha$ is the fine structure constant, $T_{BB}$ is the black-

body temperature and $n_{\mathrm{eff}}$ is the effective principal quantum number considering the quantum defects that are reported in Ref. [8]. Semi-empirical expressions for BBR-induced depopulation in high-$n$ Rydberg states ($n \gg 80$) have also been provided by Beterov et al.[9], using a quasiclassical approximation. We calculate the BB radiation induced decay rate using ARC[5] which agrees well with the semi-empirical expressions in Ref. [9]. FIG. S1

compares decay rates induced by different mechanisms at various BB temperatures, showing that 8.8% and 7% increase in the total decay rate of $30D_{5/2}$ and $28F_{7/2}$ Rydberg states respectively, is mainly due to the blackbody radiation induced decay. The rest decay channels are nearly constant at different BB temperatures. During data acquisition, radiative heating from blackbody temperatures ranging from 297.13 K to 775.13 K causes only 1.6% change in Cs gas temperature, resulting in a negligible change in atomic velocity and hence minimal impact on the transit induced decay rates. In comparison, the BB induced decay rate changes by 178.6% and 170% for $30D_{5/2}$ and $28F_{7/2}$ Rydberg states respectively in the same temperature range. In addition, collisional effects are minimized by subtracting the corresponding background signal taken at room temperature for each BB temperature.

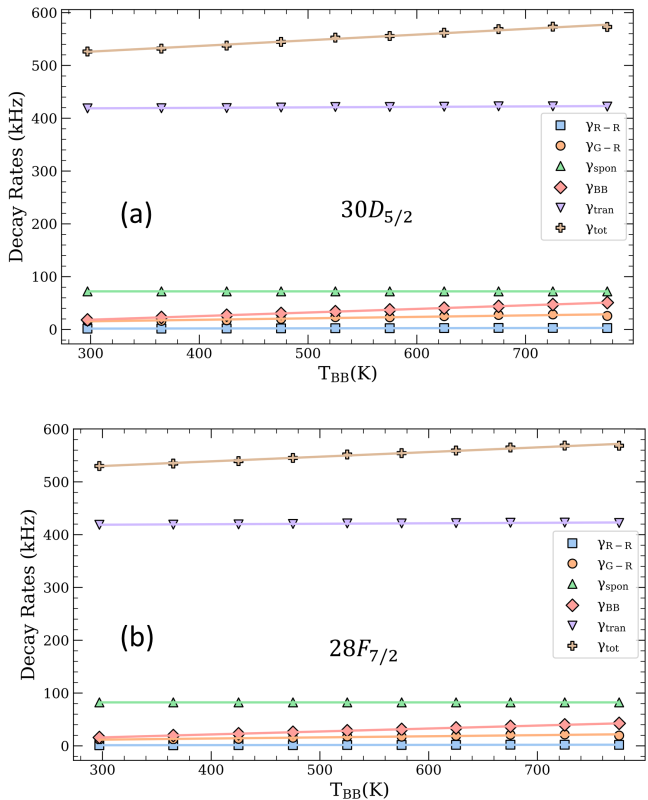

FIG. S1. Comparison of different decay rates for (a) Cs $30_{5/2}$ (b) Cs $28F_{7/2}$ Rydberg states at different BB temperatures.

\* e-mail: jshaffer@qvil.ca


\fi

\end{document}